\documentclass[aps,prl,twocolumn,groupedaddress,showpacs]{revtex4-1}
\pdfoutput=1
\usepackage{color}
\usepackage{pstool}
\usepackage{amssymb}
\usepackage{units}
\usepackage{graphicx}
\usepackage{amsmath}
\usepackage{bm}

\newcommand{\ket}[1]{| #1 \rangle}
\definecolor{matlabgreen}{rgb}{0,0.5,0}
\EndPreamble

\begin{document}

\title{Toward quantum processing in molecules: A THz-bandwidth coherent memory for light}
\author{Philip~J. Bustard}
\affiliation{National Research Council of Canada, 100 Sussex Drive, Ottawa, Ontario, K1A 0R6, Canada}

\author{Rune Lausten}
\affiliation{National Research Council of Canada, 100 Sussex Drive, Ottawa, Ontario, K1A 0R6, Canada}

\author{Duncan G.~England}
\affiliation{National Research Council of Canada, 100 Sussex Drive, Ottawa, Ontario, K1A 0R6, Canada}

\author{Benjamin~J. Sussman}
\email[email: ]{ben.sussman@nrc.ca}
\affiliation{National Research Council of Canada, 100 Sussex Drive, Ottawa, Ontario, K1A 0R6, Canada}

\date{\today}

\begin{abstract}
The unusual features of quantum mechanics are enabling the development of technologies not possible with classical physics.  These devices utilize nonclassical phenomena in the states of atoms, ions, and solid-state media as the basis for many prototypes.   Here we investigate molecular states as a distinct alternative. We demonstrate a memory for light based on storing photons in the vibrations of hydrogen molecules. The THz-bandwidth molecular memory is used to store 100-fs pulses for durations up to $\sim\unit[1]{ns}$, enabling $\sim$10$^4$ operational time bins. The results demonstrate the promise of molecules for constructing compact ultrafast quantum photonic technologies.
\end{abstract}

\maketitle
Photonic technologies increasingly utilize nonclassical physics as a fundamental aspect of device operation. Examples are wide-ranging,  from unbreakable cryptographic key generation~\cite{Bustard2011b} to single photon sources~\cite{RevSciInstrum.82.071101} and the potential speed increase offered by quantum computing~\cite{RevModPhys.79.135}. Most photonic approaches have leveraged phenomena in atoms, ions, and solid-state systems. Molecules have an array of attributes not available in these substrates, offering potential for exploitation in future quantum technology platforms~\cite{PhysRevLett.104.180501,PhysRevLett.97.033003,NJP.11.055022}. In this Letter, we demonstrate that molecules enable THz-bandwidth operation in a quantum memory, a crucial component of quantum photonic processing devices.

Unlike a classical memory, a quantum memory must preserve not a bit set as 0 or 1, but a quantum superposition state that cannot be measured in a single shot, and is susceptible to loss via decoherence~\cite{Lvovsky2009}. Memory storage lifetime is limited by decoherence caused when the storage medium interacts with its environment, resulting in loss of information. This is a general problem in the development of quantum-enhanced photonic technology; a key challenge is to identify systems in which quantum coherence is sufficiently sustained to allow device operation. The upper bound on device scale $L$ will be $L<c\tau$, where $\tau$ is the storage lifetime of the quantum memory and $c$ is the speed of light~\cite{Lvovsky2009}. For devices approaching meter length scales, nanosecond storage is necessary.

As with conventional technologies, operational speed is an important parameter for quantum-enhanced devices. In the case of memories, this can be measured in terms of maximum photon bandwidth. State-of-the-art quantum memories now operate with GHz bandwidth~\cite{Reim2010a,nature.469.512}.   Combined with the bandwidth, the decoherence time sets the number of operational time bins possible, limiting the processing potential. Much current work in photonic processing~\cite{nphoton.3.346,Science.339.794,Science.339.798} utilizes THz-bandwidth photons from pulsed parametric sources, including parametric down-conversion and spontaneous four-wave mixing. However, with typical memory bandwidths limited to the MHz-GHz regime, there have been corresponding efforts to build compatible narrow-band photon sources~\cite{nphoton.5.628}. Instead, THz-capable memories would directly enable the use of widely utilized pulsed parametric sources,  for example, to enable  synchronization of probabilistic single photons~\cite{PhysRevLett.110.133601}.

Efforts to design memories with parameters suitable for practical application have embraced a variety of substrates. Memories that interface photons to material systems consisting of a single absorbing particle, or ensemble of particles, have been developed~\cite{Lvovsky2009,Kozhekin2000,PhysRevA.65.022314}. High finesse cavities enable strong coupling of photons to single atoms~\cite{nature.473.190,PhysRevLett.98.193601}, while ensemble memories allow light--matter interaction strengths to be enhanced by storing photons in multiple weakly absorbing entities. Ensemble examples include atomic vapours~\cite{nature.452.67,nphys.5.100,Eisaman2005,nature.461.241,Liu2001,Reim2010a}, ion-doped crystals~\cite{PhysRevLett.104.080502,nature.469.508,j.lumin.127.94,nature.465.1052}, and bulk solids~\cite{science.318.1748,Lee2011a}. 

Using the unique features of molecular level spacings and matrix elements, here we demonstrate a THz-bandwidth memory, based on capturing photons in the vibrations of hydrogen molecules. Our molecular memory is used to store 100-fs pulses for durations up to $\sim$\unit[1]{ns}, enabling $\sim$10$^4$ operational time bins, demonstrating the possibility of using molecules for on-chip photonic processing devices, or compact systems approaching the meter scale.

\begin{figure*}
\begin{center}
\includegraphics[scale=0.9]{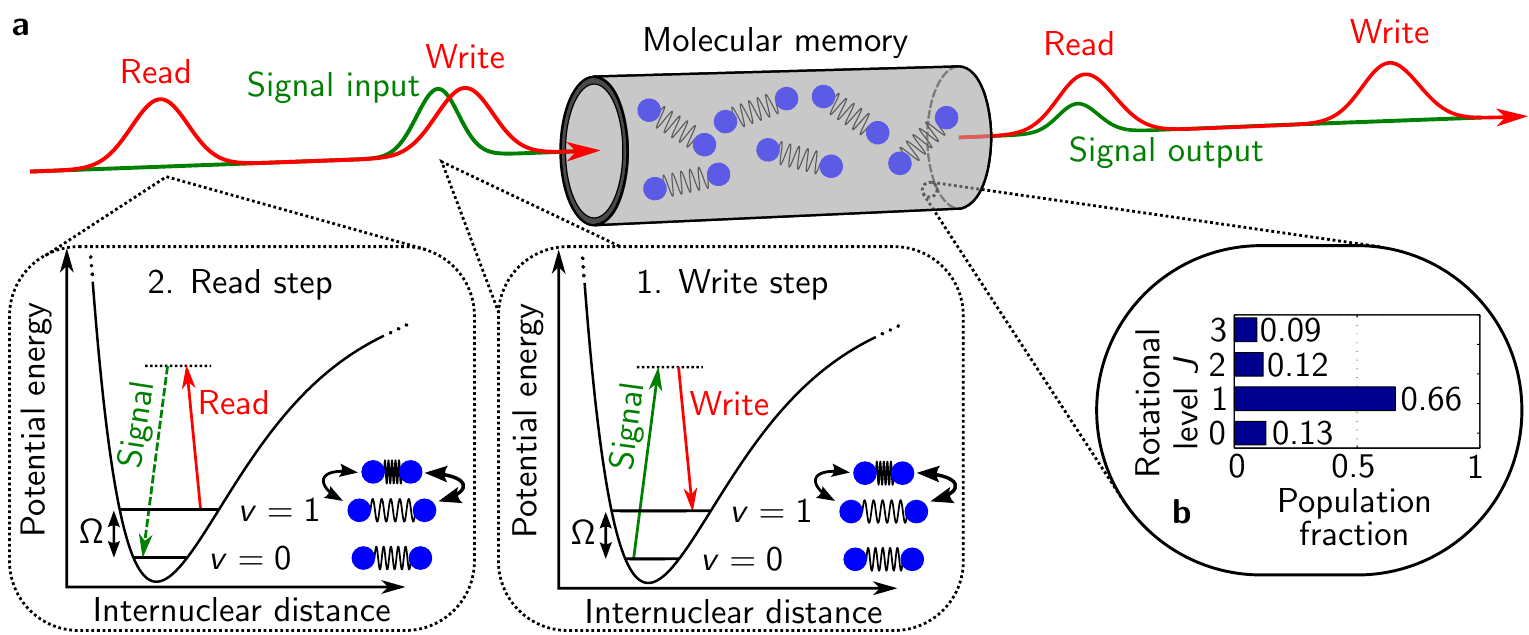}
\end{center}
\caption{(a) 1.~The write pulse induces a Stokes Raman transition where the signal pulse is absorbed in the molecular memory. 2.~After storage, the read pulse induces an anti-Stokes Raman transition where the signal pulse is emitted. (b) The room temperature population distribution among rotational levels in the $v=0$ vibrational level of hydrogen. Storage is mainly initiated from the $J=1$ level, with the other levels helping to improve the write efficiency.\label{fig:fig1}}
\end{figure*}

Our molecular memory uses  off-resonant Raman scattering~\cite{Duan2001,Kozhekin2000,Reim2010a,Nunn2007} to map photons to, and from, a vibrational excitation. The memory operation is outlined in Fig.~\ref{fig:fig1}(a) which shows an ensemble of vibrationally Raman-active molecules with all population initially in the vibrational ground state $\ket{v=0}$, and the upper storage state $\ket{v=1}$ unpopulated; all dynamics occur in the ground electronic state. In the ``write'' step, a strong write pulse of frequency $\omega_{\text{w}}$ and a weak ``signal'' pulse of frequency $\omega_{\text{s}}$ are applied in two-photon resonance with the splitting $\Omega$ between ground and storage states such that $\omega_{\text{w}}=\omega_{\text{s}}-\Omega$. This induces a Stokes Raman transition as signal photons are annihilated and mapped to  population in the storage state; the excess energy is carried away by the write pulse. After the variable storage time has elapsed, a ``read'' pulse is applied. This induces an anti-Stokes Raman transition, and the signal photons are reemitted as population is mapped from the storage state to the ground state.  

The energy level spacings of hydrogen afford a number of unique advantages for use as a memory. With a first dipole-allowed transition in the vacuum ultraviolet, optical--near-infrared fields are far detuned from resonance so THz-bandwidth photons can be stored with minimal distortion by dispersion, and negligible variation of coupling strength across the spectrum. The vibrational energy splitting of hydrogen $\Omega\approx\unit[125]{THz}$ gives a favourable Boltzmann population ratio of $2\times10^{-9}$ between the $v=0$ and $v=1$ vibrational levels at room temperature. As a result, the hydrogen sample requires no optical pumping, cooling, or other preparation for memory operation, and noise in the signal field due to anti-Stokes scattering from the thermal phonon population is low. Use of internal dynamics in neutral molecules offers additional advantages for the memory.  Since the storage excitation is a vibrational coherence with zero group velocity, the write and read processes are automatically phase matched, so interaction of the pulses is only limited by their group velocity mismatch. Furthermore, use of vibrations in neutral molecules assists in protecting the system from stray electric and magnetic fields. The information processing advantages of neutral molecules have previously been demonstrated by implementing ultrafast Fourier transforms using iodine~\cite{PhysRevLett.104.180501}. 

A 2.5-W, 1-kHz Ti:sapphire amplifier operating at $\lambda_{\text{w,r}}=\unit[800]{nm}$ generates 100-fs write/read pulses with energies of up to \unit[60]{$\mu$J} for the memory protocol. The write/read pulses are generated using an interferometer before combination with the signal pulse at a dichroic mirror. Part of the Ti:sapphire power is used to generate a signal pulse by difference-frequency generation of 1200-nm light in a superfluorescence-seeded optical parametric amplifier, followed by frequency doubling to $\lambda_{\text{s}}$=\unit[600]{nm}. The three pulses are $s$-polarized, and are focussed collinearly into a 15-cm gas cell using a 20-cm focal-length mirror. The delay of the write and read pulses relative to the signal pulse is controlled using separate optical delay lines. After the cell, light at the signal frequency is separated from the write and read pulses using spectral filters, and detected using a fast photodiode. Collisions between the molecules cause dephasing of the vibrational coherence. Unless otherwise indicated, all results presented were acquired with hydrogen pressure $p=\unit[3]{bar}$ to maximize the lifetime~\cite{PhysRevA.33.3113}.

\begin{figure}
\begin{center}
\includegraphics[scale=1]{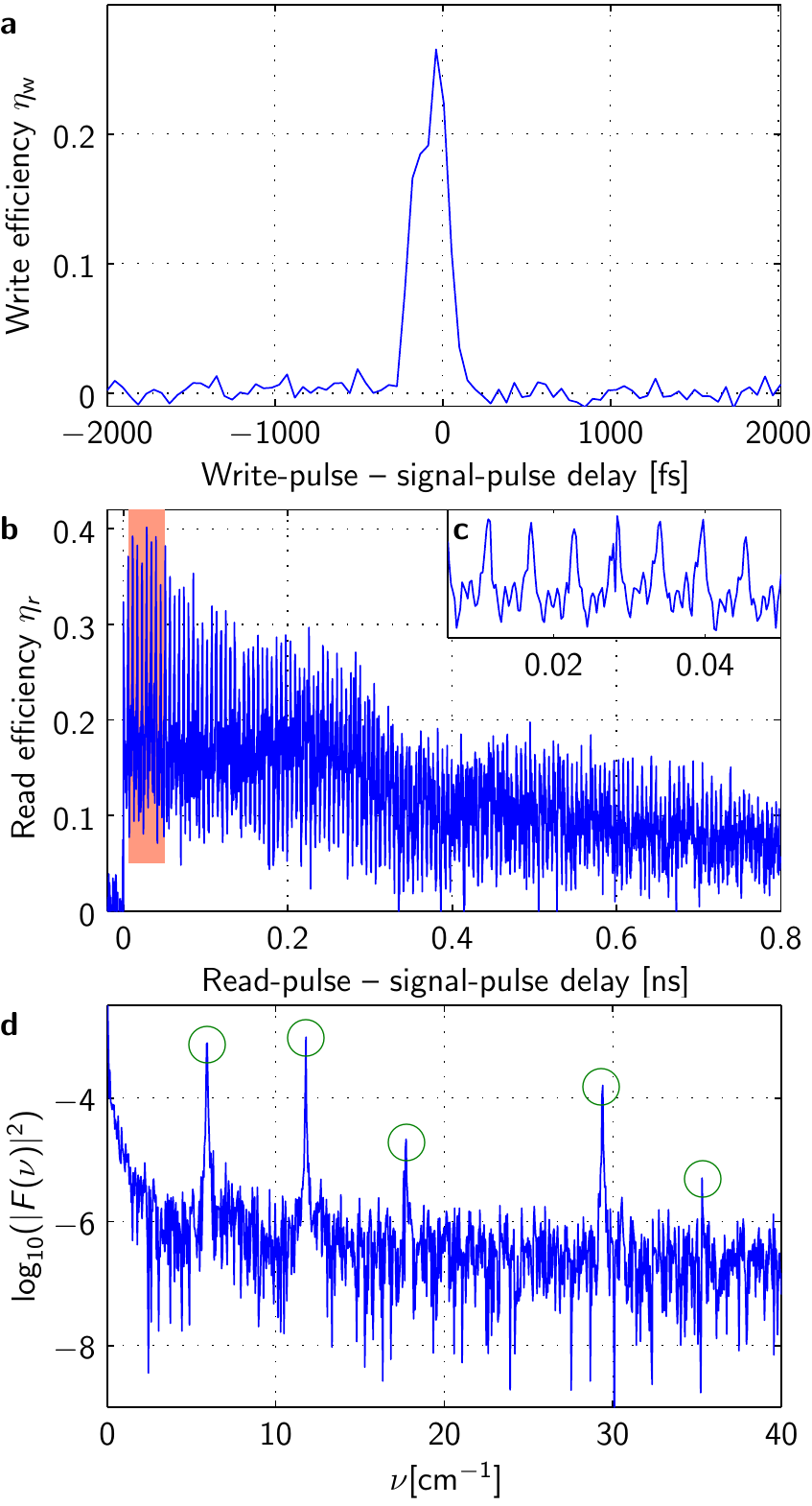}
\end{center}
\caption{(a) Plot of absolute write efficiency as a function of write-pulse -- signal-pulse delay, with no read pulse applied, showing storage of the signal pulse at $t=0$. (b) Plot of absolute read efficiency as a function of read-pulse -- signal-pulse delay. The step increase in signal at time-zero is due to re-emission of signal photons, scattered from the vibrational coherence in the molecules. The emitted signal oscillates as a function of delay because of periodic rephasing between the various rovibrational coherences. (c) High-resolution plot of shaded area from plot~(b). (d) Power spectrum $|F(\nu)|^2$ of plot~(b) on a $\log$-scale. The circled peaks, corresponding to the frequencies listed in Table~\ref{table:coherences}, show that all initially populated rovibrational levels are accessed in the memory.\label{fig:fig2}}
\end{figure}

The operation of the memory is characterized by its write efficiency $\eta_{\text{w}}$ and read efficiency $\eta_{\text{r}}$. The write efficiency is the fraction of the input signal energy that is Raman absorbed; the total efficiency $\eta_{\text{tot}}$ is the ratio of signal energy retrieved to the input signal energy. The read efficiency is then given by $\eta_{\text{r}}=\eta_{\text{tot}}/\eta_{\text{w}}$. Figure~\ref{fig:fig2}(a) shows a plot of the write efficiency as a function of write pulse delay, with the read pulse blocked.
The peak around zero delay is due to absorption into the vibrational storage state. With the write pulse delay optimized to maximize Raman absorption~\cite{Nunn2007}, Fig.~2(b) shows a plot of the measured read efficiency as a function of read pulse delay, relative to the input signal pulse. The read efficiency shows a sharp increase from zero for positive delays, due to retrieval of the stored signal field by Raman scattering. The read efficiency envelope decays as a function of read pulse delay due to collisional dephasing between the hydrogen molecules. 

The oscillations in the  read efficiency [see Fig.~2(c)] are caused by periodic rephasings of the various molecular coherences created by the write interaction. At room temperature, within the $v=0$ vibrational level, the population is distributed between the four lowest energy levels of the rotational manifold, $J=0\mbox{--}3$, where $J$ is the rotational quantum number. Nuclear spin statistics constrain the fraction of odd-$J$:even-$J$ populations to 3:1. Combined with the level spacings of hydrogen, the Boltzmann statistics deposit 66\% of the population in the $J=1$ level, with the remaining population distributed among levels with $J=0,2,3,...$ [see Fig.~\ref{fig:fig1}(b)]. The selection rules for the linearly polarized pulses used here are $\Delta v=\pm1, \Delta J=0$; vibrational transitions $\ket{v,J}\rightarrow\ket{v',J}$ are designated using the notation $Q_{vv'}(J)$. All initially populated $J$-states are within the bandwidth for two-photon coupling by the 100-fs pulses used in this experiment. As a result, all of the populated rovibrational states are used for storage of the signal pulse.  The net effect is to help increase the memory efficiency, as multiple states are used for storage. The vibrational level splittings for $Q_{01}(J)$ transitions are weakly $J$-dependent~\cite{herzberg1950spectra}. After the write interaction, each $Q_{01}(J)$ coherence evolves freely at the difference frequency between the two relevant rovibrational states. Figure~\ref{fig:fig2}(d) shows a plot of the delay-scan power spectrum, with circled spectral peaks at the beat frequencies between the different $Q_{01}(J)$ coherences which contribute to the retrieved signal field. The specific beat frequencies and corresponding transitions are listed in Table~\ref{table:coherences}. 
\begin{table}
\begin{tabular}{ l l }
\hline
$\unit[\nu]{(cm^{-1})}$ & Assigned transitions\\ 
\hline
  5.9 & $Q_{01}(0)$ and $Q_{01}(1)$ \\
  11.8 & $Q_{01}(1)$ and $Q_{01}(2)$\\
  17.6 & $Q_{01}(2)$ and $Q_{01}(3)$\\
  29.4 & $Q_{01}(1)$ and $Q_{01}(3)$\\
  35.3 & $Q_{01}(0)$ and $Q_{01}(3)$\\
\hline
\end{tabular}
\caption{Table of beat frequencies $\nu$ identified in the power spectrum of the read efficiency scan. The transitions responsible for each peak are listed in the right-hand column.\label{table:coherences}}
\end{table}
This shows that all of the initially populated rovibrational levels are accessed for storage in the memory.  Maximum readout efficiency is set by timing the readout to correspond to coherence rephasing.

Quantum memories should faithfully store the phase of an optical field. In a thermal collision-broadened ensemble memory, coherence as a function of storage time is preserved because the phase of the retrieved mode is determined by the mean ensemble phase, which is unchanged by collisions~\cite{LightScattSolids1983Shen,Belsley1993}. However, the amplitude of the retrieved pulse decays at the Raman decay rate $\Gamma$, which limits the storage time to $\sim1/\Gamma$. We demonstrated the coherence of the memory operation by interfering the retrieved signal pulse with a replica signal pulse, to observe spectral fringes. Measured in free space, the interferogram had a raw visibility of 65\%, confirming system coherence. Measuring read and write efficiencies as a function of input signal pulse energy $\mathcal{E}_{\text{s}}$ in the range $\unit[5]{nJ}\leq\mathcal{E}_{\text{s}}\leq\unit[150]{nJ}$, at a fixed control pulse delay of \unit[16]{ps}, confirms that the memory interaction is linear. Thus, due to the coherent and linear nature of the interaction, the memory is expected to be suitable for unitary operation.  

The operation of the memory is modelled by the linearized Maxwell-Bloch coupled equations for the signal field and vibrational coherence~\cite{Raymer1990}. The equations can be derived using a molecular polarizability model~\cite{Raymer1990}, or from a $\Lambda$-level formalism ~\cite{Raymer1990,Kozhekin2000,Nunn2007}. Here, the signal field is greatly detuned from the nearest optical resonance of hydrogen, and the low-lying vibrational levels are coupled by several far detuned electronic states making the polarizability approach convenient. A model of the dynamics throughout the cell requires the inclusion of coupling strength variation due to beam propagation. However, due to the nonlinearity of the Raman coupling, storage is strongly weighted around the focus. Therefore, in this geometry, the dynamics may be described using a one-dimensional model~\cite{Raymer1990}. The memory efficiency is then determined  by the dimensionless coupling parameter $G=g(\omega_{\text{s}})I_{\text{w}}z\Gamma\tau_{\text{w}}$ where $g(\omega_{\text{s}})$ is the steady-state Raman absorption coefficient, $I_{\text{w}}$ is the write pulse intensity, $z$ is the propagation distance, $\Gamma$ is the collisional damping constant, and $\tau_{\text{w}}$ is the write-pulse duration.  

\begin{figure}
\begin{center}
\includegraphics[scale=1]{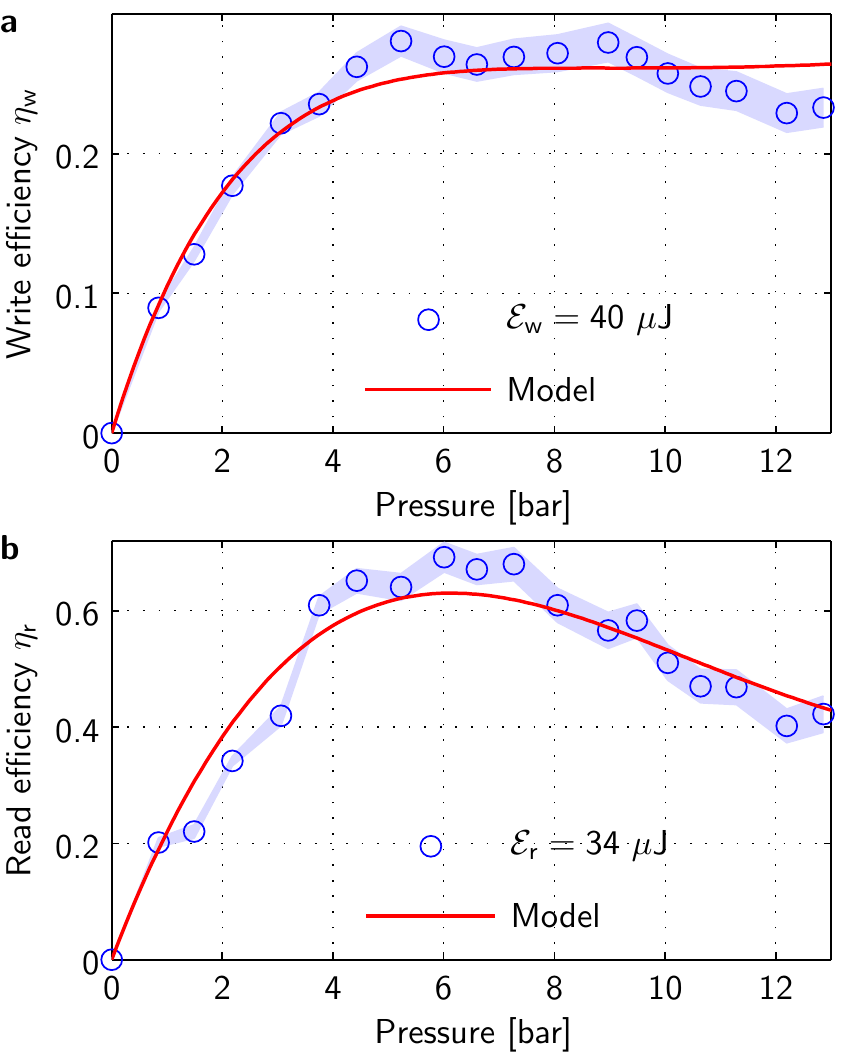}
\end{center}
\caption{(a) Plot of write efficiency as a function of gas pressure for write pulse energy $\mathcal{E}_{\text{w}}=\unit[40]{\mu J}$ (blue circles). (b) Plot of read efficiency as a function of gas pressure for read pulse energy $\mathcal{E}_{\text{r}}=\unit[34]{\mu J}$ (blue circles). Increasing the density increases the storage cross section, but can also result in reabsorption of the emitted field so that there is an optimal pressure around \unit[6]{bar}. Standard errors on the measurements are shown in the shaded regions. Simulations of the measured efficiencies are plotted in the solid red curve.\label{fig:fig3}}
\end{figure}

Increasing the gas pressure increases the product $g\Gamma$, which is linearly proportional to number density, allowing the memory operation to be optimized. The measured read and write efficiencies are plotted as a function of gas pressure in Fig.~\ref{fig:fig3}.
With the read pulse blocked, Fig.~\ref{fig:fig3}(a) shows the measured write efficiency as a function of gas pressure for a write pulse energy of $\mathcal{E}_{\text{w}}=\unit[40]{\mu J}$ (blue circles).  With the read-pulse--write-pulse delay set to a rephasing maximum at \unit[16]{ps}, the read efficiency is shown in Fig.~\ref{fig:fig3}(b) as a function of gas pressure for read pulse energy $\mathcal{E}_{\text{r}}=\unit[34]{\mu J}$ (blue circles). Figure~\ref{fig:fig3} also shows theoretical simulations (solid red curves) of the efficiencies based on numerical solution of the Maxwell-Bloch equations. A mode-mismatch term $\alpha$ is included in the simulations of the write efficiency to account for the imperfect spatial, temporal, and spectral matching of the input signal pulse with the write pulse. The fit indicates that $\alpha=35\%$ of the signal pulse is mode matched for writing to the memory. No mode mismatch is included for the read efficiency simulation because the write interaction only stores photons which are well-matched with the write and read pulses. The simulation curves show broad qualitative and quantitative agreement with the measured efficiencies. The read efficiency is larger than the write efficiency due to the imperfect mode matching for the write interaction. At a pressure of $p=\unit[13]{bar}$ we achieve a maximum value of $G\simeq6.5$, extracted from the theoretical model. Increasing the pressure to this level increases the interaction strength, but simultaneously increases the possibility of reabsorption of the retrieved signal. These counteracting effects result in an optimum pressure for maximum efficiency at $p=\unit[6]{bar}$.

The maximum total efficiency of the memory demonstrated here was $\eta_{\text{tot}}=18\%$. Accounting for the mode mismatch parameter, the effective total efficiency is $\eta_{\text{tot}}/\alpha\sim51\%$. Improved spectral, temporal, and spatial mode matching should therefore allow us to increase the total efficiency. It may also be possible to achieve high-efficiency operation with lower write/read pulse energies by using hollow-core capillaries~\cite{PhysRevA.65.041802} or hollow-core photonic crystal fibers~\cite{science_298_399} to extend the interaction region, which is presently limited by free-space diffraction. Furthermore, fiber integration of the molecular memory will enable photonic processing with compact, on-chip devices.

We have demonstrated a THz-bandwidth, tunable, photonic memory for femtosecond pulses.  Using Raman transitions, we have mapped photons into the vibrations of molecules, and retrieved them at a variable delay, up to \unit[1]{ns}.   The lifetime of the memory could  be enhanced by cooling the hydrogen sample~\cite{PhysRevA.33.3113}, or by using vibrons in solid parahydrogen~\cite{PhysRevA.58.R58}.  Good agreement is obtained between the experimental data and a Maxwell-Bloch model.  This memory demonstrates the viability of using molecules for quantum photonic processing. We expect such a memory to be suitable for local photonic processing in on-chip devices or bench-scale systems.

\acknowledgements{The authors thank Josh Nunn for helpful comments on the manuscript.}

\end{document}